\documentstyle[12pt,aasms4]{article}
\newcommand{\rg}{R_g}

\newcommand{\kes}{\kappa_{es}}

\begin{document}

\title{Hot Atmospheres Around Accreting Neutron Stars:\\
A Possible Source For Hard X--ray Emission}
\author{Silvia Zane\altaffilmark{1}}
\affil{International School for Advanced Studies, \\ Via Beirut 2-4, 34014
Trieste, Italy \\ e--mail: zane@sissa.it}
\altaffiltext{1}{also at: Dept. of Physics, University of Illinois
at Urbana--Champaign, 1110 W. Green St., Urbana, IL 61801--3080, USA} 
\author{Roberto Turolla}
\affil{Dept. of Physics, University of Padova, \\ Via Marzolo 8, 35131 Padova,
Italy \\ e--mail: turolla@astaxp.pd.infn.it}
\and
\author{Aldo Treves\altaffilmark{2}}
\affil{International School for Advanced Studies, \\ Via Beirut 2-4, 34014
Trieste, Italy \\ e--mail: treves@astmiu.uni.mi.astro.it}
\altaffiltext{2}{present address: II Faculty of Sciences, University of 
Milano, Via Lucini 3, 22100, Como, Italy}

\begin{abstract}
The structure of static atmospheres around
unmagnetized neutron stars undergoing steady, spherical accretion is discussed.
We focus on the ``hot'' configurations presented by \markcite{tur94}Turolla
et al. (1994) and calculate the radiation spectrum using a 
characteristics method. In particular, it is found that e$^+$--e$^-$ pair
production may affect significantly  the external atmospheric layers, where 
positron and proton number densities become of the same order. The consequent
increase of the scattering opacity lowers the Eddington limit and this, in
turn, may drive a dynamical instability if the accretion luminosity is large
enough, ultimately producing a rapid expulsion of the envelope. If ``hot''
states are indeed accessible, this mechanism could give rise to transient
phenomena in hard X--rays of potential great astrophysical interest. 
\end{abstract}


\section{Introduction}

The problem of calculating the radiation spectrum produced in the atmosphere 
of an 
unmagnetized, accreting neutron star (NS) received wide attention in the past, 
starting from the pioneering work by \markcite{zs69}Zel'dovich, \& Shakura 
(1969, hereafter ZS) where the case of 
stationary, spherical accretion was first discussed. In the commonly 
accepted picture, steaming from ZS' original analysis and corroborated 
by the more detailed numerical work by \markcite{al73}Alme, \& Wilson 
(1973), the emerging spectrum is essentially a  blackbody plus a 
high--energy Compton tail. It can be characterized by two parameters, the 
accretion luminosity $L_{\infty}$ (or, equivalently, the accretion rate) and
the column density $y_0$ corresponding to the penetration range of incoming
protons in the NS atmosphere. Recently \markcite{zam95}Zampieri et al. 
(1995, paper II)
extended ZS's results to cover the low luminosities typical of old, isolated
NSs accreting the interstellar medium ($L\lesssim 10^{34}$ erg/s), 
showing that deviations from 
a blackbody become more conspicuous as the luminosity decreases. 

\markcite{tur94}Turolla et al. (1994, paper I) 
re--examined the issue of the thermal and radiative properties of
static atmospheres around accreting NSs and pointed out 
that a new class of much hotter equilibrium solutions may exist if the
accretion luminosity exceeds a certain critical value, in  analogy
with what has been already found in black hole accretion (see e.g.
\markcite{par90}Park 1990; \markcite{ntz91}Nobili, Turolla, \& Zampieri 
1991). The existence of two
states is related to the different role thermal and non--thermal processes
may play in exchanging energy between photons and electrons in the
atmosphere: in the ``cold'' state (\`a la ZS) the energy released by accretion
is radiated away essentially via bremsstrahlung, whereas in the ``hot''
state Compton scattering dominates (in this respect see also 
\markcite{kat80}Burger, \& Katz 1980). 

In paper I ``hot'' solutions were found 
solving the frequency--integrated transfer problem. The lack of
any information about the angular and frequency dependence of the
radiation field made impossible to evaluate the photon--photon pair 
production, which was nevertheless recognized to
be potentially important, and led to introduce rather drastic assumptions in
the calculation of the Compton energy exchange rate.

In this paper we try to overcome some of these limitations, in the 
attempt to shed light on the relevance of ``hot'' solutions.
To get some physical insight without resorting to a full time--dependent
calculation, we consider the stationary case and analyze the effects of 
different processes separately. First the radiation field is 
calculated using the characteristics method introduced by Zane et al. (1996, 
paper III).
Pair production is neglected and the frequency--dependent transfer
equation is solved coupled to the hydrostatic and energy balance.
The resulting intensity is then used to estimate the pair number
density $n_+$ produced by photon--photon interactions. Although  
not fully self-consistent, 
this approach shows that a large pair production is
expected in the external atmospheric layers. This opens the possibility that
such configurations are unstable with respect to the onset of a
radiatively--driven wind, produced by the increased opacity in the pair--rich
plasma. Astrophysical applications of ``hot'' atmospheres 
are suggested in the context of hard X--ray transient phenomena. 

\section{The Model}

In this section we briefly outline the input physics of our model, referring
to previous works (ZS; \markcite{al73}Alme, \& Wilson 1973; paper I and II) for
all details, and introduce our radiative transfer calculations.

We consider an
unmagnetized, non--rotating neutron star with mass $M_* = 1 M_\odot$ and radius
$R_*=10^6$ cm, surrounded by a static atmosphere 
where the accreting flow is decelerated by Coulomb collisions and/or plasma 
interactions. As the flow hits the outermost stellar layers, the proton bulk 
kinetic energy is converted into electron thermal energy and then 
re--emitted as electromagnetic radiation. The heat 
injected per unit time and mass, $W_h$, is given by
(see \markcite{bsw92}Bildsten, Salpeter, \& Wasserman 1992; paper II)
\begin{equation}
W_h = {L_{\infty} \over 8 \pi R_*^2 y_0 y_g } { 1 +  v_{th}^2 /v_i^2  
\over \left [ 1 - \left ( 1 - v_{th}^4 / v_i^4 \right ) \left ( y / y_0 
\right ) \right ]^{1/2}} 
\end{equation}
where $y_g = \left ( 1 - \rg /R_* \right )^{1/2} $ is 
the gravitational redshift factor, $\rg = 2GM_* / c^2$, $v_{th}^2 = 3KT 
/m_p$, $T$ is the gas temperature, $v_i^2 = c^2 \left ( 1 - L_{\infty}/L_E 
\right )(\rg/R_*)$, $y = \int_R^{\infty} \rho dR$ is the column density 
and $L_E$ is the Eddington luminosity.   
As discussed in paper I, these solutions show no significant
envelope expansion, so the radial coordinate can be safely assumed equal to 
the star radius. The atmospheric structure can be solved 
in plane--parallel geometry,  using as independent variable the 
scattering optical depth $\tau = \kes y$, where $\kes$ is the Thomson 
opacity. In this case, the luminosity profile can be
immediately derived from the zero--th frequency--integrated moment equation (see
paper II).
The run of pressure $P$ and temperature $T$ are given by the solution of 
the hydrostatic and energy balance equations (see again paper I and II)
\begin{equation}
{{dP}\over{d\tau}} = {G M_* \over 
y_g^2 R_*^2 \kes } \left ( 1 - 
y_g{ \kappa_1 \over \kes} {L\over{L_E}}  \right 
) \, , 
\end{equation}
\begin{equation}
{ W \over c}  = \kappa_P \left ( a 
T^4 - { \kappa_0 \over \kappa_P} U \right 
) + (\Gamma -\Lambda)_C \, ,  
\end{equation}
with the boundary condition of vanishing pressure at the top of the atmosphere.
In equation (3) $U$ is the radiation energy density, 
$\kappa_P$, $\kappa_0$ and $\kappa_1$ are the Planck, 
absorption, and flux mean opacities and $(\Gamma - \Lambda)_C$ is the Compton 
energy exchange rate.
The effective heating $W$ represents the fraction of 
$W_h$ which is converted into electromagnetic radiation within the atmosphere; 
$W$ coincides with $W_h$ if no energy is injected at the inner boundary,
that is if $L_{in} = 0$. Since at large depths the radiation field 
becomes more and more isotropic, $L_{in}$ is indeed small, although in a 
realistic situation it is not exactly zero. On the other hand, the thermal 
structure of both the ``cold'' and ``hot'' solutions presented in paper I 
proved not sensitive to the assumed value of the luminosity at the bottom of 
the atmosphere, provided it is not too large compared with $L_\infty$. 
In particular, ``hot'' atmospheres exhibit a rather sharp drop in the gas 
temperature which separates two 
distinct regions characterized by different thermal properties. 
In the external layers, where the bulk of the luminosity is produced, the 
energy released by accretion is radiated away via Compton cooling, while 
in the deeper, denser layers LTE is attained and the temperature profile 
is mainly determined by bremsstrahlung equilibrium.
For these solutions $L_{in}$ also coincides with the value of the 
radiative flux at the top of the cold region. 

There are two important points concerning the ``hot'' models that 
were not thoroughly discussed in paper I and deserves further 
analysis. Both of them concern the possibility of establishing and maintaining 
a rather
steep temperature gradient between the hot to the cold part of the atmosphere,
so that a ``hot'' solution can indeed set in.
First of all we note that the position of the temperature drop is practically
unrelated 
the value of the proton stopping length. Although for 
$y_0 = 20$ (the value we will use here) they practically coincide, we stress 
that, in general, the value of 
the scattering depth at which the transition occurs depends only on the 
relative efficiency of Compton and bremsstrahlung heating--cooling and, as 
numerical tests show, it is always close to $\tau_{tr} = 10$ ($y_{tr} =
25$). 
The presence of a sudden decrease in $T$ can be understood comparing the 
free--free and Compton thermal times
\begin{equation}
{{t_C}\over{t_{ff}}} = {{\epsilon_{ff}\left(1 - U/aT^4\right)}\over
{4k|T-T_\gamma|\kappa_{es}U/m_ec^2}} \, ,
\end{equation}
where $\epsilon_{ff}$ is the free--free emission and $T_\gamma$ is the 
radiation temperature.
As the scattering depth increases $T$ and $T_\gamma$ become closer, 
because non--conservative scatterings tend to establish thermal 
equilibrium, and $U$ either increases or stays constant. 
In the hot, effectively thin layers $U\ll aT^4$ and at large enough $\tau$, 
where $T\simeq T_\gamma$, the time scale for free--free cooling becomes 
shorter than the Compton time. The plasma temperature drops
until $aT^4 \simeq U$ in order for the energy balance to be satisfied 
and the radiation temperature follows. 

The second important point concerns the efficiency of electron conduction
at the interface between the hot and cold regions. This effect was not
considered in paper I. In order to get a quantitative
estimate of the relevance of thermoconduction we compare the energy
flux due to conduction, $F_c = \nu_c\rho dT/dy$, with the radiative flux, 
$F_r = L/4\pi R^2$. Using the expression for $\nu_c$ given in 
\markcite{zr67}Zel'dovich, \& Raizer (1967) it is
\begin{equation}
{{F_c}\over{F_r}} \sim 2.5\times 10^{-3}\left({T\over{10^9}}\right)^{5/2}
\left({L\over{L_E}}\right)^{-1}{y_{tr}\over{\Delta y}}
\end{equation}
where $\Delta y$ is the width of the transition layer and $T$ is the 
temperature in the hot region.
As the previous equation shows, electron
conduction is indeed efficient in transferring heat from the hot to the
cold region. However, if a fraction of the luminosity is produced below
$y_{tr}$, $L_{in}/L_{\infty} \sim 0.2-0.3$ for $L_\infty\sim 0.1 L_E$, 
conduction 
will limit the temperature of the hot region to $\sim 10^9$ K, smear the
jump over $\Delta y\sim 2-3$ and then cease to be important. The stationary
temperature profile will then look quite similar to that discussed by 
\markcite{zr67}Zel'dovich, \& Raizer (1967) in the context of shock waves in 
a plasma. Although the real profile will deviate from that of
paper I (mainly in the absence of the peak at $T\sim 10^{10}$ K, see curve b
in figure 2, which will be leveled if conduction is taken into account), the 
emerging picture seems to be substantially unaltered by the inclusion of
thermal conduction. This conclusion is reinforced by the result of
the frequency--dependent calculation presented here 
which shows that the temperature profile is nearly isothermal at 
$T\sim 10^9$ K (see curve a in figure 2). 
It is interesting to point out that 
the existence of a ``hot'' equilibrium state can be recovered 
analytically by solving the energy balance equation in the limit in which 
Compton cooling dominates and using an approximated expression for
the Comptonized radiation spectrum (see \markcite{tit97}Titarchuk 1997, 
\markcite{tlm97}Titarchuk, Lapidus \& 
Muslimov 1997). In the case in which Comptonization is effective and produces
a rather flat spectrum with $\alpha\ll 1$ and the scattering depth at
the stopping radius exceeds unity it is $T\sim 10^9$ K quite independently
on the value of $y_0$ (Titarchuk, private communication).

In the following we consider in more detail ``hot'' atmospheres. 
As we discussed, in this picture only a small fraction of the total luminosity 
($L_{in}/L_{\infty}\lesssim 0.2-0.3$) is produced below $y_{tr}$, where the 
scattering depth is already larger than $\sim 10$. 
Bearing this in mind, we can compute the radiation field only for 
$y< y_{tr}$ with a suitable boundary condition at $y_{tr}$. The consistency
of the computed flux at the boundary with the assumed value of $L_{in}$
will be checked a posteriori.
Radiative transfer is solved by means of 
the characteristics method discussed in paper III
and yields the full spatial, angular and frequency dependence of 
the photon occupation number $f= c^2 I/2h^4 \nu^3 $, where $I$ is
the specific intensity.
In the case of a static, plane--parallel atmosphere $f$ depends on 
one ``spatial'' variable, $\tau$, on the cosine of the angle between the 
photon propagation direction and the vertical direction, $\mu$,
and on the local photon energy $E = h \nu$. The spectrum observed at infinity
is obtained redshifting to energies $E_\infty = E (1-\rg/R_*)^{1/2}$
that  at the top of the atmosphere. We used as boundary conditions $f=0$ at 
the top of the atmosphere for incoming rays and a regularity 
condition ($df/d\tau=0$) at the inner boundary for outgoing rays. The 
interested reader is referred again to paper III 
for all computational details about the source term which includes
relativistic e--p, e--e bremsstrahlung and Compton scattering.
In particular, Compton scattering is treated in its more general form 
by direct evaluation of the integrals of the Compton scattering kernel.
To avoid numerical instabilities which arise when more accurate 
formulae are used (see e.g. \markcite{skp88}Shestakov, Kershaw, \& Prasad 
1988), the Kompaneets approximation was retained in the calculation of the 
Compton energy exchange rate. 

The resulting monochromatic energy density is shown in figure 1 for 
$L_{\infty}/L_E = 7 \times 10^{-2}$, $L_{in}=0.2L_\infty$ and $y_0 = 20$ 
g cm$^{-2}$ at different optical depths.  The dashed 
line represents the spectrum observed at infinity, i.e. corrected for the 
gravitational redshift. This model has been 
computed solving the transfer equation for 20 values of $\mu$, 25 energies 
in the range 0.017 MeV $\leq E\leq 5$ MeV and $10^{-3}\leq\tau\leq 8$. 
The considered luminosity is close to the lower limit
for the existence of ``hot'' solutions with $y_0=20$ g cm$^{-2}$, and it can 
be thought to be well representative of situations where photon energies 
are large enough to produce pairs, since $T_\gamma$ is anticorrelated with
$L_\infty$ (see paper I). The emergent flux, in fact, is peaked at 
$\sim 500$ keV and is characterized by a high--energy tail.
The gas and 
radiation temperatures, together with the correspondent profiles derived 
in paper I, are shown in figure 2. The lower values of $T$ (about a factor 3)
are mainly due to the different calculation of $T_{\gamma}$ which enters in 
the energy equation. Now the radiation temperature
is directly evaluated from the mean intensity and not obtained from 
a phenomenological equation, as in paper I. In the present calculation, the 
atmosphere appears  to be nearly isothermal. In both cases 
the energy balance is practically obtained  equating 
Compton heating and cooling, which gives $T\sim T_\gamma$. 
Now $T$ is always close to $10^9$ K and the 
computed luminosity at the inner boundary is $\sim 0.02 L_E\sim 30\% L_\infty$
(quite close to the input value), so conduction is not expected to be much 
effective.

\section {Pair Processes}

We use the results derived in the previous section to estimate the
pair number density $n_+$, under the simplifying hypothesis that pairs 
thermalize with electrons and protons in the atmosphere. Clearly this is
just an {\it ad hoc\/} assumption that needs to be checked a posteriori
(see the discussion at the end of the section).

Let us consider the process 
\begin{equation}
\gamma + \gamma \rightleftharpoons e^+ + e^-\, ;
\end{equation}
quantities 
relative to the two photons will be denoted with indices 1, 2. 
The absorption coefficient for a photon of dimensionless 
energy $x_1=h\nu_1/m_ec^2$ 
propagating in direction $\mu_1$ through a radiation field of intensity
$I_2 = I (\tau, x_2, \mu_2)$ is given by 
\begin{equation}
a_{\gamma \gamma}(x_1) = { 1 \over hc } \int {I_2 \over x_2} \sigma 
\left ( x \right ) \left ( 1 - \mu \right ) dx_2 d \mu_2 d \phi_2 
\, , 
\end{equation}
where $x^2 = x_1 x_2 \left ( 1 - \mu \right ) /2 $ is the square of the photon 
energy in the center--of--momentum frame, $\mu$ is the
cosine of the angle between the two photon directions, $\phi_2 $ is 
the azimuthal angle and 
$\sigma (x)$ is the photon--photon pair production cross--section (see 
e.g. \markcite{jr76}Jauch, \& Rohrlich 1976). Replacing 
$\phi_2$ with $\phi = \phi_2 - \phi_1$ and introducing $x_\pm^2 = x_1 x_2 
\left [ 1 - \cos \left ( \theta_1 \pm \theta_2 \right ) \right ] /2$ (see
\markcite{sg83}Stepney, \& Guilbert 1983), expression (7) becomes
\begin{equation}
a_{\gamma \gamma} = { 4 \over hc } \int d x_2 d \mu_2\, {I_2 \over x_1 x_2^2  } 
F \left ( x_+ , x_- \right ) \, .
\end{equation}
The function  
\begin{equation}
F \left ( x_+ , x_- \right ) = \int_{x_-}^{x_+} 
{ \sigma 
\left ( x \right ) x^3 dx \over 
\sqrt { \left ( x_+^2 - x^2 \right ) \left ( x^2 - x_-^2 
\right ) }}  
\end{equation}
does not depend on the specific intensity so it was calculated once for all
on a fixed grid of values of $x_\pm$ and stored in a matrix. 
The outer double integral was then computed using a spline interpolation.    
Once the absorption coefficient is known, the pair production rate 
$R_{\gamma \gamma}$ is obtained performing two further integrations 
over angles and energies
\begin{equation}
R_{\gamma \gamma} = { \pi \over h } \int dx_1 d\mu_1\, {I_1 \over x_1} 
a_{\gamma \gamma}  \, .  
\end{equation}

Following \markcite{sv82}Svensson (1982) and introducing $\theta = KT/m_ec^2$, 
the pair annihilation rate can be expressed in terms of a 
dimensionless function $A(\theta )$ as 
\begin{equation}
\dot n_+ = n_+ n_- c r_e^2 A \left ( \theta \right ) \, ,
\end{equation}
where 
\begin{equation}
A \left ( \theta \right ) \simeq { \pi \over 1 + 2 \theta^2 /  
\ln \left ( 2 \eta \theta \right ) }\, ,
\end{equation}
$\eta\simeq 0.56146$ and $r_e$ is the classical electron radius. Combining 
the previous expressions, the pair balance equation becomes in the stationary 
case
\begin{equation}
R_{\gamma \gamma} - 
z ( 1 + z ) n^2 c r_e^2 A \left ( \theta \right ) = 0 \, , 
\end{equation}
where $z = n_+/n$, $n$ is the proton number density and the charge neutrality 
condition $n_- = n_+ + n$ has been used. 
The positive root of equation (13) is 
\begin{equation}
z(\tau ) = { 1 \over 4 } \left ( \sqrt { 1 + { 16 R_{\gamma \gamma} \over n^2 
A(\theta )r_e^2 c^2 }} - 1 \right ) \, .  
\end{equation}
Using the specific intensity and the run of thermodynamical variables 
derived in section 2, we calculated the proton and positron number densities 
shown in figure 3, together with $z$. As it can be seen, while in the inner 
atmospheric regions no relevant pair production is expected, the value of 
$z$ becomes $\sim 10$ for $\tau \approx 10^{-3}$ and tends to increase in 
the external regions. 

Since the Eddington limit in a pair plasma is  lowered by a factor $m_e/m_p$,
there is the possibility that large values of $z$ are never reached, 
because pairs are accelerated outwards by the radiative force as soon as 
they are created. The characteristic timescale for radiative acceleration
$t_{acc}\sim (2h/a)^{1/2}\sim 5\times 10^{-9} 
(h/10)^{1/2}(L_\infty/L_E)^{-1/2}$ s,
$h\sim 10$ cm is the scale height of the envelope and $a\sim (GM/R_*)(L/L_E)$
is the radiative acceleration, should be compared 
with the timescales for particle collisions and for the development of 
plasma instabilities. Coulomb collisions establish thermal equilibrium in a 
time $t_{coll}\sim 6\times 10^{-7}(T/10^9)^{3/2}(n/10^{18})^{-1}$ s, while
the growth time for the counter--streaming instability is $t_{ins}\sim 2\times 
10^{-11} (n/10^{18})^{-1/2}$ s (see e.g. \markcite{mel86}Melrose 1986). 
Plasma instabilities have therefore  enough time to produce microturbulence
which, in turn, will efficiently couple $e^\pm$ to the ambient plasma. 
Enhanced particle scattering prevents pairs from escaping and also allows
them to thermalize with atmospheric electrons. Our initial assumption
that pair production/annihilation is in equilibrium seems indeed justified.

\section{Discussion and Conclusions}

The existence of ``hot'' solutions, first proposed in paper I, has been
confirmed by means of a more detailed calculation of the radiation field.
Our frequency-- and angle--dependent approach allowed us
to compute the specific intensity  and hence to evaluate the photon--photon 
pair
production. The temperature and density profiles are close to those presented 
in paper I and the spectrum peaks at about $500$ keV.

First of all, we wish to comment about the possibility of
getting a ``hot'' solution started. Of course, the thermal energy
stored in a ``hot'' atmosphere is much larger than in a ``cold'' one, and 
it is roughly 
\begin{equation}
U_{th}\sim 4\pi R_*^2 y_{tr}\left({{KT}\over{m_p}}\right)\, , 
\end{equation}
where $T$ is the average temperature. Assuming $KT\sim 100$ keV, $U_{th}\sim 
10^{31}$ erg, a factor $\sim 100$ above the ``cold'' case. 
The transfer of energy from protons to electrons in our
model is described by equation (1). Obviously, this is a rather crude estimate
which does not enter into the details of the physical processes responsible
for the energy exchange. Heating associated with the proton stopping 
might produce temperatures only in the keV range, so the transition between 
``cold'' and ``hot'' states could require some more efficient heating 
process, like dissipation of shock waves or magnetic field reconnection. 
We note, however, that the proton bulk kinetic energy at the neutron star
surface is $\sim 100$ MeV, which is much higher than the temperature of 
``hot'' solutions. In any case, the smallness of the energy content of the 
atmosphere compared to the value of the luminosity, $L_\infty\sim 10^{37}$ 
erg/s, strongly indicates that there should be no severe physical
hindrance to drive the transition on time scales of microseconds.

On the other hand, one can argue that  the ``hot'' state 
should be short--lived. In fact,
as shown is section 3, pair production is expected to become important,
at least in the outer layers where the pair density reaches its equilibrium
value at $z\sim 1-10$. Due to the extra opacity produced by $e^\pm$, the 
critical luminosity in the atmosphere becomes a factor $1+2z$ lower than the 
Eddington limit. In ``hot'' models with $L_\infty\sim
0.1 L_E$, where pairs appear to be coupled with the plasma, part of the 
envelope will become necessary dynamically unstable 
when $z\sim 5$. This conclusion probably remains valid even if the appearance
of pairs alters the temperature profile and in a complete self--consistent
solution the value of $z$ is lower than that considered here. 
One should expect the outer layers to be expelled very quickly at the onset
of the ``hot'' state: as a consequence accretion may be inhibited 
with the possible production of a relativistic shock wave. 
This suggestion is corroborated by the consideration that
the momentum fluxes of the two flows are within an order of magnitude, 
as can be seen comparing $\dot M_{acc}v_{acc}$ and 
$\dot M_{out}v_{out}\approx M_{out}a$, where $M_{out}\sim 4\pi R_*^2
y$ is the mass in the unstable part of the atmosphere and 
$a\sim GM[(1+2z)L/L_E -1]/R_*^2$ is the acceleration. Using the 
numerical results of section 3
and assuming $\dot M_{acc}\sim L(R_*/GM)$, $v_{acc}= c(\rg/R_*)^{1/2}$, which
gives an upper limit for the momentum flux of the ingoing material, it turns 
out that  $\dot M_{acc}v_{acc}\lesssim 10\dot M_{out}v_{out}$.
The decrease of the accretion rate and of the luminosity may push the system
in a regime where only the ``cold'' solution exists, giving rise to an 
on--off behaviour.

The ``hot'' state should be characterized by a spectrum close to that shown in
figure 1, with typical emission at $\approx 100$ keV.  Moreover, pair bursts
are expected, each consisting of $\sim  4 \pi R_*^2yz/m_p \sim 6\times 10^{35}$
particles; an upper limit for the integrated luminosity in the annihilation
line will be of order of $10^{29}$ erg. These hot, optically thin and 
Comptonized envelopes bear some resemblance with hot coronae above accretion
disks, although the geometry, the heating process, and the triggering of the 
pair instability may be different. The onset of the ``hot'' state
represents a possible physical mechanism for producing efficiently high energy
radiation from weakly magnetized, accreting neutron stars and may be of interest
in connection with hard X--ray transients, although we are aware that
some simplifying assumptions, like spherical symmetry and absence of
magnetic fields, may limit the applicability of our results.
 
At present hard emission ($E\gtrsim 30$ keV) has been observed 
from a number of sources, 
believed to contain NSs, with SIGMA and BATSE (see e.g. 
\markcite{tl96}Tavani, \& Liang 1996; \markcite{var96}Vargas et al. 
1996). At least 
seven X--ray bursters, including Aql X--1, and other LMXBs show 
transient emission in the range 30--200 keV, with
intensity anticorrelated with that in the soft band. 
These observations are of the utmost importance, since hard X--rays emission
were previously associated only with X--ray pulsators and black hole candidates.
Tavani, \& Liang have proposed a possible explanation for the transient, hard
state of these sources in terms of magnetic reconnection in the inner regions
of an accretion disc around the NS, but the nature of the primary emission
mechanism is still an open issue. 
The total X--ray luminosity of these sources is $\sim 10^{36}-10^{37}$ erg/s, 
close to that considered in our model, although the duration of the hard
state is several days. In our picture, this would imply that the appearance 
of e$^\pm$ pairs, even if it affects the accretion process, does not induce the 
transition to the ``cold'' state. Hard states of shorter duration
would be impossible to detect with BATSE and SIGMA, but they may be observed 
by the instrumentation on board of XTE or SAX.

\acknowledgments

We are indebted both to an anonymous referee for drawing our attention on
the importance of thermal conduction in ``hot'' models and to the second
referee, Lev Titarchuk, for his clarifying comments. The revised 
version of this paper greatly benefited from discussions with 
Fred Lamb, Luciano Nobili and Luca Zampieri.
We are also grateful to Marco Tavani for useful discussions about hard X--ray
transients.

\clearpage

\figcaption[fig1.eps]{Monochromatic mean intensity for the model 
$L_{\infty} = 7 \times 10^{-2}L_E$, $y_0 = 20$ g cm$^{-2}$; different lines
correspond to equally spaced values of $\log\tau$ in the interval $[-3, 0.9]$. 
The emerging redshifted spectrum is also shown 
(dashed line).\label{fig1}}

\figcaption[fig2.eps]{a) The gas temperature (full line) and the radiation 
temperature (dashed line) for the model of figure 1. b) Same results from 
the frequency--integrated analysis  of paper I.\label{fig2}} 

\figcaption[fig3.eps]{Proton (full line), positron (dashed line) number 
densities (in units of $10^{22}$ cm$^{-3}$) and $z=n_+/n$ (dash--dotted line) 
versus the scattering optical depth for the model of figure 1.\label{fig3}} 


\clearpage

\begin{references}

\reference{al73} Alme, M.L., \& Wilson, J.R. 1973, \apj, 186, 1015
\reference{kat80} Burger, H.L., \& Katz, J.I. 1980, \apj, 236, 921
\reference{bsw92} Bildsten, L., Salpeter, E.E, \& Wasserman, I. 1992,
\apj, 384, 143
\reference{jr73} Jauch, J.M., \&  Rohrlich, F. 1976, 
The Theory of Photons and Electrons (New York: Springer--Verlag)
\reference{mel86} Melrose, D.B. 1986, Instabilities in Space and Laboratory 
Plasmas (Cambridge: Cambridge University Press)
\reference{ntz91} Nobili, L., Turolla, R., \& Zampieri, L. 1991, \apj, 383, 250
\reference{par90} Park, M--G. 1990, \apj, 354, 64
\reference{skp88} Shestakov, A.I., Kershaw, D.S., \& Prasad, M.K. 1988,
J. Quantit. Spectros. Radiat. Transfer, 40, 577
\reference{sg83} Stepney, S., \&  Guilbert, P.W. 1983, \mnras, 204, 1269
\reference{sv82} Svensson, R., 1982, \apj, 258, 335
\reference{tl96} Tavani, M., \& Liang, E. 1996, \aap, in press
\reference{tit97} Titarchuk, L. 1997, in proceedings of the 2nd Integral 
Workshop, ESA SP-382
\reference{tlm97} Titarchuk, L, Lapidus, I., \& Muslimov, A. 1997, \apj,
submitted
\reference{tur94}Turolla, R., Zampieri, L., Colpi, M., \& Treves, A. 1994, 
\apj, 426, L35 (paper I)
\reference{var96} Vargas, M., et al. 1996, \aap, 313, 828
\reference{zam95} Zampieri, L., Turolla R., Zane S., \& Treves, A. 1995, \apj, 
439, 849 (paper II)
\reference{zan96} Zane S., Turolla, R., Nobili, L., \& Erna, M. 1996, 
\apj, 466, 871 (paper III)
\reference{zr67} Zel'dovich, Ya, \& Raizer, Yu. 1967, Physics of Shock Waves
and High--Temperature Hydrodynamic Phenomena (New York: Academic Press)
\reference{zs69} Zel'dovich, Ya., \& Shakura, N. 1969, Soviet Astron.--AJ, 
13, 175 (ZS)

\end{references}
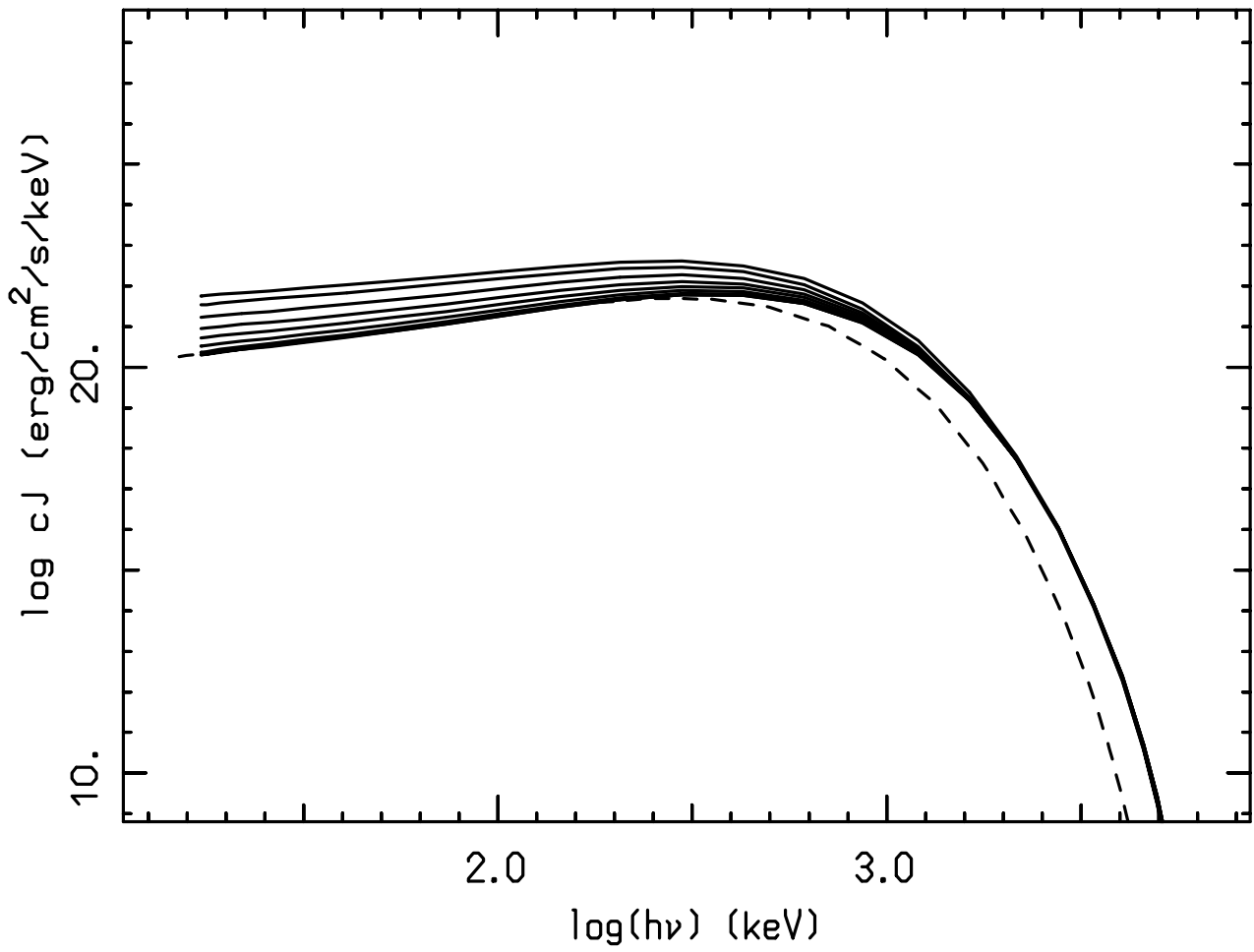
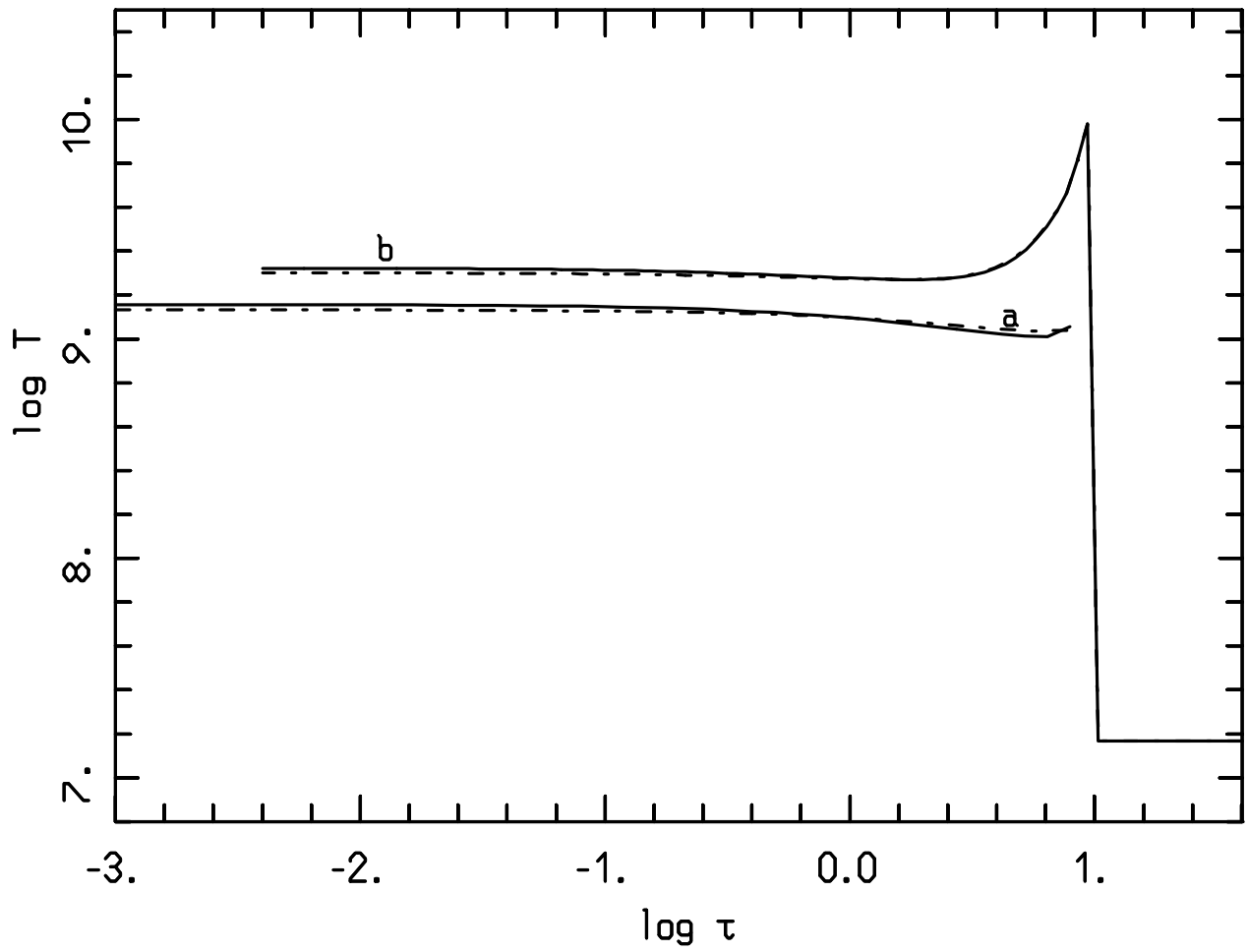
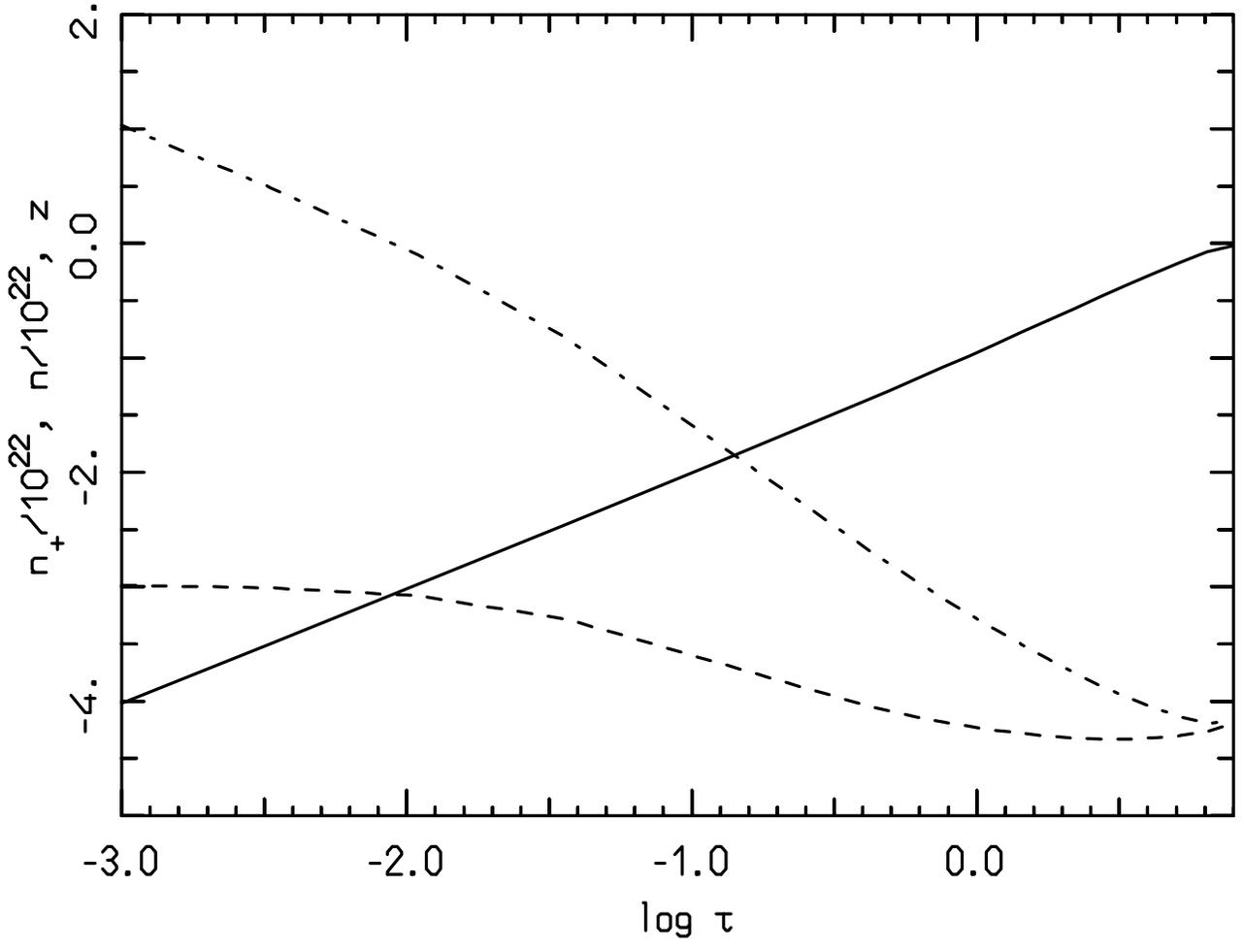
\end{document}